\documentclass{aa}
\usepackage{epsfig}
\begin{document}
\def\h50{h$_{50}^{-1}${}}
\def\kms{km~s$^{-1}${}}
\thesaurus{11.01.2; 11.06.2; 11.09.2; 11.16.1; 11.17.3; 13.09.1}

\title{Near-infrared photometry of isolated spirals with and without
an AGN.  I: The Data.  \thanks{Based on data obtained at: the European
Southern Observatory, La Silla, Chile, the T\'elescope Bernard Lyot,
Calar Alto Observatory, Las Campanas Observatory. Also based on
observations made with the NASA/ESA Hubble Space Telescope, obtained
from the data archive at the Space Telescope Institute.}}

\author{
  I.~M\'arquez \inst{1,2}
\and
  F.~Durret \inst{2,3}
\and
  R.M.~Gonz\'alez Delgado  \inst{1}
\and
  I.~Marrero \inst{1}
\and
  J.~Masegosa \inst{1}
\and
  J.~Maza \inst{4}
\and
  M.~Moles \inst{5}
\and
  E. P\'erez \inst{1}
\and
  M. Roth \inst{6}}
\offprints{I. M\'arquez (\sl{isabel@iaa.es}) }
\institute{
    Instituto de Astrof\'\i sica de Andaluc\'\i a (CSIC),
Apartado 3004 , E-18080 Granada, Spain
\and
    Institut d'Astrophysique de Paris, CNRS, 98bis Bd Arago,
F-75014 Paris, France
\and
    DAEC, Observatoire de Paris, Universit\'e Paris VII, CNRS (UA 173),
F-92195 Meudon Cedex, France
\and
    Departamento de Astronom\'\i a, Universidad de Chile, Casilla 36D, 
Santiago, Chile
\and
    Instituto de Matem\'aticas y F\'\i sica Fundamental (CSIC),
  Madrid, Spain and Observatorio Astron\'omico Nacional, Madrid, Spain. 
  Presently on sabbatical leave at Queen Mary \& Westfield College,
London, UK
\and
    Observatories of the Carnegie Institution of Washington, 813 Barbara Street, Pasadena, CA91101
}
\date{Received,  ; accepted,}

\maketitle

\begin{abstract}

We present infrared imaging data in the J and K' bands obtained for 18
active spiral galaxies, together with 11 non active galaxies taken as
a control sample. All of them were chosen to satisfy well defined
isolation criteria so that the observed properties are not related to
gravitational interaction. For each object we give: the image in the
K' band, the sharp-divided image (obtained by dividing the observed
image by a filtered one), the difference image (obtained by
subtracting a model to the observed one), the color J-K' image, the
ellipticity and position angle profiles, the surface brightness
profiles in J and K', their fits by bulge+disk models and the color
gradient.

We have found that four (one) active (control) galaxies previously
classified as non-barred turn out to have bars when observed in the
near-infrared. One of these four galaxies (UGC 1395) also harbours a
secondary bar. For 15 (9 active, 6 control) out of 24 (14 active, 10
control) of the optically classified barred galaxies (SB or SX) we
find that a secondary bar (or a disk, a lense or an elongated ring) is
present.

The work presented here is part of a large program (DEGAS) aimed
at finding whether there are differences between active and non active
galaxies in the properties of their central regions that could be
connected with the onset of nuclear activity.

\keywords{galaxies: active - galaxies: fundamental parameters -
galaxies: photometry -
infrared: galaxies }
\end{abstract}

\section{Introduction}

The existence of non-axysimmetric components of the galactic potential
has been frequently invoked as an efficient way to transport gas from
the galaxy scale down to the nucleus to fuel the AGN. In particular,
the shocks and gravitational torques induced by a galactic bar 
\footnote{we mean bar $=$ primary bar; when considering secondary bars, it 
will be clearly stated} can
make the gas loose angular momentum and therefore facilitate the
fuelling mechanism (Shlosman et al. 1989). Nevertheless, it has been
suggested that there is not a preference for Seyfert nuclei to occur
in barred galaxies (Heckman 1980; Simkim et al. 1980). This result 
is confirmed by statistical analyses on the catalogued morphology of
Seyfert galaxies (Moles et al. 1995) and through near-infrared imaging
stu\-dies (McLeod \& Rieke 1995; Alonso-Herrero et al. 1996; Mulchaey \&
Regan 1997).

We have started a large observational program aimed at studying the
conditions for the onset of activity in galactic nuclei. Based on
previous work (Moles et al. 1995), we search for detailed
morphological and kinematical differences between active and non
active galaxies of similar global morphology. We particularly pay
attention to those that could facilitate the transport of gas towards
the very central regions and the nucleus (for an example of the
complete analysis of one of our sample galaxies see P\'erez et
al. 1999). For this purpose, we are obtaining optical and
near-infrared images and long slit spectroscopy with the best possible
spatial resolution.

We selected a sample of galaxies, out of which 18 host active nuclei
and the rest form a control sample of non active galaxies 
(see Sect. \ref{sample}). 

In this paper we present the first data set: the infrared imaging data
in the J and K' bands.  Infrared imaging is particularly important
because it allows to trace the old stellar population, and to separate
the various components (the bulge, disk, bar(s) and spiral arms) with
the smallest contribution of the active nucleus and less contamination
by dust absorption. In fact, various K band imaging studies have
revealed that bars were present in galaxies classified as unbarred in
the optical (McLeod \& Rieke 1995; Mulchaey \& Regan 1997), hence
showing that the near-infrared is better suited for these
purposes. The sample is described in Section 2. In Section \ref{data}
we present the near-infrared data we have obtained together with the
HST archive images we used. In Section \ref{methods} we describe the
different methods of analysis applied to the images. In Sections
\ref{actives} and \ref{non-active} the results are described for each
of the galaxies in both sets of active and non-active spirals
respectively. The summary is given in Section~2.

A discussion on the NIR properties of these galaxies, together with
the comparison of active versus non-active galaxies derived from these
data will be presented
in a companion paper (M\'arquez et al., in preparation).

\section{The sample }

The active galaxies have been chosen with the following criteria:
(a)~Seyfert 1 or 2 from the V\'eron-Cetty \& V\'eron (1993) catalogue;
(b)~with morphological information in the RC3 Catalogue; (c)~isolated,
in the sense of not having a companion within 0.4 Mpc (H$_0$=75
km/s/Mpc) and cz$<$500 km/s, or companions catalogued by Nilson
without known z; (d)~nearby, cz$<$6000 km/s; and (e)~intermediate
inclination (30 to 65$^\circ$). The control sample galaxies have been
selected among spirals verifying the same conditions (b), (c), (d) and
(e), and with types similar to those of the active
spirals. Thus, all the galaxies in our sample are isolated, in the
sense of avoiding possible effects of interactions with luminous
nearby galaxies. \footnote{in a strict sense, they are weakly
interacting systems, since we don't take into account the possibility
of being surrounded by dwarf galaxies or HI clouds.} 

The main properties of the sample galaxies, as given in the RC3
catalogue are presented in Table \ref{sample}. 

The samples are not large enough to allow statistically significant
results, but the detailed study of each of the galaxies in these
samples should give us at least hints on what differences between
active and non-active isolated spiral galaxies, if any, are to be
taken into account to identify the conditions for the onset of nuclear
activity.

The active sample extends farther in redshift. In particular, there
are six active galaxies with cz$>$ 4000 km/s, but none in the control
sample. Although this changes the physical resolution achieved for
those objects, we don't expect any bias due to that difference.
Absolute mean B luminosities give M$_{\rm B}=-20.4 \pm 0.6$ for
Seyferts and M$_{\rm B}=-20.3 \pm 0.9$ for non-active galaxies. Some
contribution from the active galaxies would be expected as a
consequence of the presence of the active nucleus itself and of the
higher levels of star formation sometimes observed in that kind of
galaxies (Maiolino \& Rieke 1995). If any, this effect is not
systematically present in our small samples, which can therefore be
considered as comparable with respect to the host galaxy luminosity.
The distribution of inclinations is almost flat for the non-active
galaxies, whereas there is a deficiency of active galaxies at large
inclinations, as expected for an optically selected sample of active
galaxies (McLeod \& Rieke 1995). The morphological types of spiral
active galaxies are earlier than Sbc, resembling the type distribution
of active galaxies in the V\'eron-Cetty \& V\'eron catalogue (Moles et
al.  1995); non-active galaxies have been selected to have similar
types. Looking at the presence of bars in the galaxies of both samples
(as catalogued in the RC3), we notice that only one galaxy in the list
of non-active objects is unbarred, and three more are of intermediate
type X. For the active galaxies, half are barred and four more of type
X. As already quoted, the presence of a primary bar does not seem to
be the direct cause for the nuclear activity. Otherwise the lists of
active and non active galaxies we have observed are rather similar in
global properties.  We therefore estimate that the sample of
non-active galaxies is well suited to be used as a control sample.

\section{The data }\label{data}

We have obtained K' images for the 18 active galaxies together with J
images for 15 of them.  J and K' images were taken for 11 non-active
spirals. The infrared data have been acquired mainly with three
telescopes: the 2m T\'elescope Bernard Lyot (Pic du Midi Observatory,
France), the 2.5m DuPont telescope (Las Campanas, Chile), and the 3.5m
telescope at Calar Alto (Spain). We have also used some short
observations taken with the ESO 2.2m telescope in La Silla for the
relative calibration of one of the J images obtained during a
non-photometric night. The journal of observations is presented in
Table \ref{journal}.  We always used NICMOS-3 256$\times$256
detectors; the pixel scales are given in Table \ref{journal}.

As seen in Table \ref{sample}, there are some
galaxies with extended angular sizes close to 2 arcmin or even larger,
which do not fit in a single frame.  For two of them, namely NGC 4779
and NGC 6951, a mosaic of images was performed in order to have images
for the entire galaxy disk.

The data reduction and calibration was performed, following the
standard procedures with the IRAF\footnote{IRAF is the Image Analysis
and Reduction Facitily made available to the astronomical community by
the National Optical Astronomy Observatories, which are operated by
the Association of Universities for Research in Astronomy (AURA),
Inc., under contract with the U.S. National Science Foundation.}
software and SQIID package.  For the objects, exposures of typically
about 2 minutes (dithered in short exposures of about 20s in J and 50s
in K' offset by about 2-3 arcsec) were taken on the source and on the
sky, until the total integration time was achieved (see Table
\ref{journal}).  A median dark frame was calculated for every night
with the same exposure time than that of the objects by using the
corresponding dark frames obtained before and after the observations.
Flat-field frames were obtained for each filter by subtracting the
median dark frame from the sky frames, normalizing the resulting
frames and calculating the median. We then subtracted the dark frame
from the object frame, divided by the normalized flat-field and
subtracted the corresponding sky frame. Sometimes, another constant
had to be subtracted in order to attain the zero background level.  In
order to align the resulting frames to better than a pixel fraction we
used the non-saturated stars in these frames or, when there were no
stars, we used the centre of the galaxy. For the flux calibration a
number of standard stars were observed and mean extinction
coefficients were applied (k$_J$ and k$_{K'}$ in magnitudes
airmass$^{-1}$ are respectively 0.12 and 0.08 for Las Campanas and La
Silla, 0.12 and 0.10 for the TBL, 0.057 and 0.042 for the NOT and
k$_{K'}$=0.1 for Calar Alto).  The error level on the standard stars
amount to 10\%. In column 7 of Table \ref{journal} we give the
isophotal magnitude/arcsec$^2$ corresponding to 2$\sigma$ of the
background.

We have retrieved HST calibrated infrared images obtained with NICMOS
(filter F160W) for the eight galaxies for which such data was
available (see column 1 in Table \ref{journal}).

\section{Methods of analysis}\label{methods}

\subsection{Morphological features}

K' images of the galaxies in grey scale with isocontours overlaid are
shown if in Figs. 1a-28a. All figures are only available in electronic
form.

In order to detect the presence of features showing a departure from
radial symmetry such as bars and spiral arms, we have used a masking
technique. We filtered the original images with a box of 2-3 times the
FWHM of the seeing (median filter) and divided the observed image by
the filtered one.  The resulting images (hereafter called the
sharp-divided images to differenciate them from those obtained by
subtraction, i.e. the so-called unsharp masking technique) are shown
in Figs. 1b-28b.  This technique is very well suited to trace
asymmetries in the light distribution, such as bars, spiral arms, dust
lanes, rings; it allows the subtraction of the diffuse background in a
very convenient way to look for subtle, small-scale variations and
discuss the possible presence of both dust extinguished and more
luminous regions (Sofue et al. 1994; M\'arquez \& Moles 1996;
M\'arquez et al. 1996; Erwin \& Sparke 1999; Laine et al. 1999).

The presence of bars can be determined quantitatively by studying the
behaviour of isophotal position angles (PAs) and
ellipticities ($\epsilon$s):
a bar is characterized by a local maximum in the
$\epsilon$ corresponding to a constant PA (Wozniak et al. 1995;
Friedli et al. 1996; Jungwiert et al. 1997). Therefore, we have fit
ellipses to the isophotes (Jedrzejewski 1987) with the IRAF ellipse task in
stsdas.analysis.isophote, allowing the center, PA and
$\epsilon$ to vary from one isophote to the next. Foreground stars
have been masked out. The resulting parameters allow the reconstruction of a 
model galaxy. The difference between
the original image and the model (hereafter called the di\-fference
images) are particularly useful to trace all the features that cannot
be described with ellipses, as spiral arms and boxy
or peanut-like components. They are given in Figs. 1c-28c and
discussed separately for each galaxy.
When the PA suddenly changes from one isophote to the
next, the model cannot work for the regions left inside (see Figs.
1c, 8c, 9c, 11c, 12c, 15c, 18c, 19c, 25c and 26c).

\subsection{Surface photometry and photometric decomposition}

The major axis PA (counted anticlockwise from north to east as usual)
and $\epsilon$ from the ellipse fitting are drawn as a function of
the ellipse semi-major axis (therefore, in radius) in
Figs. 1e-28e. 

In Figs. 1f-28f the surface brightness profiles (isophotal magnitude
fit to the ellipse {\it versus} ellipse semi-major axis) are given for K'
 and J. We have used the 1-D resulting
profiles to obtain the bulge and disk contributions. We have fit
an exponential to the outermost region (the disk),
subtracted it to the observed profile and fit a r$^{1/4}$ law (the
bulge) to the residual; this calculation was continued until
convergence was achieved (see M\'arquez \& Moles 1996, 1999). The
resulting bulge and disk parameters are given in Table
\ref{decomposition}. The surface brightness profiles in J and K' are
plotted in Figs. 1f-28f with the bulge, disk and bulge+disk fits
superimposed.  The corresponding residuals are shown in
Figs. 1g-28g. Since residual background subtraction is the biggest source of
error in determining the profiles (see for instance de Jong 1996), we
caution the reader that in the cases where the galaxy occupies most of
the frame and the residual 
background level could only be determined in very small
regions, the error in the profiles can reach 15-20\%. Otherwise,
residuals are in general smaller than 20\%, except in the bar and
spiral arm regions.

\subsection{Color images and color gradients}

For galaxies for which a J band image is also available, we display in
Figs. 1d-28d the ratio of the J to K' images (after background
subtraction), median filtered by 3$\times$3 pixels in order to enhance
the S/N of the outer regions. Color images will allow a more complete
description of the previously detected morphological features in terms
of the contribution of dust and/or star forming events. We note that
in the cases where the background is not flat, somewhat structured
color images are obtained (galaxies with non-flat remaining
backgrounds are indicated with an asterisk in column 1 of Table
\ref{journal}).

Magnitudes integrated in circular apertures and J-K' color gradients
were obtained from the curves of growth as in M\'arquez \& Moles
(1996). The apertures and the corresponding magnitudes are given in
Table \ref{decomposition}. Color gradients are shown in Figs. 1h-28h
for galaxies for which we have data in the two bands. The background
subtraction was done as described above and the same caveat applies,
in the sense that some color gradient could be artificially produced
for the galaxies with non-flat backgrounds.

In Table \ref{compara} we give the magnitudes that we measure for 9
galaxies, together with J, K' and/or K magnitudes found in the
literature.  A straightforward comparison between our J values and
those reported in previous works (for the same apertures) gives:
J$_{us} -$ J$_{other} = -0.42 \pm 0.25$. Only 4 galaxies in our sample
have been observed in K' by Mulchaey et al. (1997); we find: K'$_{us}
-$ K'$_{\rm Mulchaey} = 0.015 \pm 0.63$. If we exclude NGC 6890 (see
section 5.17) we obtain: K'$_{us} -$ K'$_{\rm Mulchaey} = -0.30 \pm
0.36$. For 10 galaxies and a total of 12 measurements, we find:
K'$_{us} -$ K$_{other} = -0.42 \pm 0.32$. This value is comparable to
the difference between K'$_{\rm Mulchaey}$ and K$_{other}$ ($-0.55 \pm
0.49$).  These differences in the zero points are not critical
for our purposes and will not be discussed further. In any case, we
notice that the differences in the colour indexes J-K' are within the
errors.

In Table \ref{bars} we give the parameters we determine for primary
and secondary bars \footnote{we note that all the elongated central
structures we call {\it secondary bars} 
could be either bars or elongated rings or disks;
kinematical data will help to discriminate among them.}  in the sample
galaxies. We specify what of the methods above have been used to
detect the presence of secondary bars (or primary bars when detected
for the first time), and if evidences are found in available HST
images.  We note that, excepting the secondary bar of IC~2510, only
seen in the sharp-divided image (just marginally in the PA-$\epsilon$
plot), the secondary bars are obtained in at least two methods.

\section{Description of individual active galaxies.}
\label{actives}

\subsection{UGC~1395}

The K' image in Fig. 1a shows an extended bulge from which two spiral
arms originate.  A small elongation is seen in the center, that could
be due to the presence of a small bar (see below). Although this
galaxy is classified as S(rs)b in the RC3, McLeod \& Rieke (1995)
already reported the presence of a bar with a 16 arcsec radius,
PA=145\degr~ and $\epsilon$=0.5 from their K image. It doesn't show up
in our sharp-divided image in Fig. 1b, but is clearly detected in the
plot of $\epsilon$ and PA with radius (Fig. 1e) which is in agreement
with Peletier et al. (1999). The parameters we deduce for the bar from
that plot are very similar to those reported by McLeod \& Rieke (1995).

The sharp-divided image reveals the presence of a small bar, extending
along PA = 139\degr, up to 7 arcsec from the center. It is not
detected in the PA and $\epsilon$ plot as it is too weak to be
apparent.  Other evidence for this nuclear bar is the curved dust
pattern that surrounds the very central region in the broad band HST
image (filter F606W) by Malkan et al.  (1998).

The difference image in Fig. 1c shows that the overall fit is good
except in the region where the arm contribution is important.  The
surface brightness profile (in agreement with that of McLeod \& Rieke)
is well fit by a bulge$+$disk model, except in the zone where the arm
contribution is important (Figs. 1f and 1g).

\subsection{IC~184}

This galaxy shows rotation of the PA and twisting of the isophotes in
the central region.  Evidence for the existence of a bar inside the
primary bar is presented in Figs. 2b and 2e, where $\epsilon$ shows
two maxima for a rather constant PA.  The inner bar is also evident in
the broad band HST image by Malkan et al. (1998). The region
connecting the two bars is visible in Fig. 2b as a thin curved
elongation starting at the end of the inner (thicker) bar.

The difference image in Fig. 2c shows the two nested bars as well as
the region where the spiral arms begin; the spiral arm to the north is
much brighter than its southern counterpart. Due to the bars and arms,
the residuals are high except in the very outer zones (Fig. 2g), and
the bulge + disk fit is not very good (Fig. 2f).

\subsection{IC~1816}

A big bar (Fig.  3e) and three spiral arms are detected, among
which the north west arm is the brightest and most detached
(Fig. 3a). At large scales, the image has a somewhat triangular shape.
A small galaxy is seen 35 arcsec to the south along PA=168\degr, but no
redshift is available
for it.

The sharp-divided image (Fig. 3b) clearly reveals the bar, 
the ring, and the spiral arms. Note
that the very central zone appears somewhat elongated.

The bar and arms are also clearly seen in the difference image (Fig.
3c), and except for the bar region the residuals are quite small
(Fig. 3g). The bulge + disk model fits the data very
well except for the bar region (Fig. 3f).

The J/K' image (Fig. 3d) has a rather smooth aspect except at the very
center where a small, elongated structure is seen, just in the
direction traced by the offset dust lane in the broad band HST image
by Malkan et al. (1998). It could be the signature of a nuclear bar within
a 1 arcsecond radius (see also Fig. 3e).

\subsection{UGC~3223}

This is a classical barred SBa galaxy, with a very large bar 
(Figs. 4a and 4e). The sharp-divided
image (Fig. 4b) clearly reveals the bar as well as the beginning of
two spiral arms originating at both extremities of the bar.

The bulge $+$ disk model fits well the observed profile except in the
bar region (Fig. 4f), and except for this (large) region the residuals
are very small (Fig. 4g).  The spiral arms appear most clearly in the
difference image (Fig. 4c).

We find evidence for the presence of an inner bar in Fig. 4e (see
Table \ref{bars}).  In fact, an inner bar feature is evident in the
HST broad band image by Malkan et al. (1998), with a curved dust lane
structure associated with it.

\subsection{NGC~2639}

This galaxy appears to be very regular at first sight (Fig. 5a).
However, the sharp-divided image (Fig. 5b) reveals the presence of a
fat primary bar, at variance with its RSa(r) classification. We note
that Fig. 5e shows a roughly constant PA value for the whole galaxy,
precluding a definitive quantification of the primary bar
extension. In any case, NGC 2639 has to be classified as barred (or
lensed) on the basis of the IR images.

The difference image reveals the existence of two spiral arms, one to
the north west and the other to the south east, which splits into two
arms (Fig. 5c). The surface brightness profile (in agreement with that
by Moriondo et al. 1998) decomposition, specially in the K' band
(Fig. 5f), implies very small residuals except in the lense and the
spiral arm region (Fig. 5g).

A red arc-like feature going from the SE spiral arm northwards to the
NW arm is visible in Fig. 5d, corresponding to the presence of dust
lanes which clearly appear in the broad band HST image by Malkan et
al. (1998).  This would indicate the presence of a small central
bar. An elongated ring-like structure from 9 to 18 arcseconds in
radius is seen in the Pa$\alpha$ HST image by B\"oker et al. (1999),
i.e., a ring just exterior to the bar. This small bar is further
confirmed by the sharp-dividing method and the ellipse fitting applied
to the HST H image (see Figs. 5e and 29).

\subsection{IC~2510}

The K' band image shows a bar, spiral arms and a pseudo-ring
(Fig. 6a). The spiral arm to the south 
is clearly visible on the sharp-divided image (Fig.  6b), as well as
a secondary bar.

The difference image (Fig. 6c) shows very distinctly the spiral. 
This is the only case for which we detect a secondary bar only from one method.
Moreover, primary and secondary bars have almost the same PA.
We therefore quote the secondary bar as doubtful in Table \ref{bars}.

The bulge $+$ disk decomposition gives good results
except in the bar region, specially in K' (Figs. 6f and 6g).  The J/K'
image remains quite constant (Figs. 6d and 6h).

\subsection{NGC~3281}

The K' image (Fig. 7a) shows an elongated boxy bulge with possible
spiral arms coming out of it. 
This galaxy is classified as non-barred in the RC3. However, the
sharp-divided image (Fig. 7b) clearly shows the existence of a weak
bar. The possible
existence of a bar was already suggested by Xanthopoulos (1996), on
the basis of an I band image. We also detect weak spiral arms in the
sharp-divided image. Since these structures are weak, the PA is
roughly constant throughout the galaxy (Fig. 7e); however, they do
induce a small bump in the surface brightness profiles (Fig.  7f).

The J/K' image (Fig. 7d) shows that the (J-K') color index is fainter in
the nucleus, in the
small spiral structure around the nucleus, and in the region of the
bar.

\subsection{NGC~3660}

A strong bar is seen in the K' image of this galaxy (Fig. 8a) but the
ring is barely visible.  On the other hand, both the bar and ring
appear clearly on the sharp-divided image (Fig. 8b).

The edges of the bar can also be seen in the difference image
(Fig. 8c); the bar parameters are determined from Fig. 8e. 
>From an R image, Chapelon et al. (1999) give PA=116\degr~
out to 16 arcsec for the bar. This agrees with Friedli
et al.  (1996) who show that bars are generally longer in K than in R.

The J/K' color image (Fig. 8d) shows a smooth structure, increasingly
red towards the center, and a steep central gradient (Fig. 8h). The
bar region also appears redder than the surroundings. The color map
also shows clear hints of the existence of a small, red circumnuclear
region.

Due to the large size of this galaxy, 
the bulge+disk decomposition is not reliable
(Fig. 8f), as confirmed by the high residuals (Fig. 8g).

\subsection{NGC~4253}

This galaxy has a thick bar and a weak external ring (Fig. 9a). Its
nucleus is displaced relatively to the centroid of the outer
isophotes. The bar parameters from Fig. 9e are in agreement with
previous results in J (Alonso-Herrero et al. 1998) and K (McLeod \&
Rieke 1995; Peletier et al. 1999).  Note that this direction is also
that of the stellar bar seen at optical wavelengths
(Mulchaey \& Wilson 1996, Mulchaey et al. 1996).

The sharp-divided image (Fig. 9b) reveals the presence of a small
structure, possibly a secondary bar, roughly perpendicular to the main
bar. This feature appears
even more clearly on the difference image (Fig. 9c).

The J/K' image shows a double nuclear structure with a (J-K') color
index redder than the rest of the galaxy (Fig. 9d); this
structure can explain the observed decentering. 
Notice that in the very central zone ($r$ $<$ 3 arcsec) 
$\epsilon$ is not the same in J and K
(Fig. 9e). This feature seems to correspond to that delineated by the
dust pattern that surrounds the innermost 2 arcsec in the HST optical
image (Malkan et al. 1998). Unfortunately, the nucleus is saturated in
the infrared HST image, so we cannot analyse the
presence of faint nuclear elongations. 
Nevertheless, a curved dust pattern feature with a radius of
$\approx$4 arcsec can be seen along the NE-SW direction (Fig.  30),
favouring the presence of a nuclear bar.

The profile is in good agreement with that given by McLeod \& Rieke
(1995). The bulge+disk fit is quite satisfactory (Figs. 9f and
9g). 

\subsection{NGC~4507}

The K' image agrees with that of Mulchaey et al. (1997), and indeed shows
no clear presence of a bar
(Fig. 10a). A noticeable curved dust lane
to the SE reaching the innermost 2 arcsec is visible in the HST image
by Malkan et al. (1998).

The sharp-divided image (Fig. 10b) clearly shows a
bar that is also
apparent on the difference image (Fig. 10c). This bar had been already
reported by Mulchaey et al. (1997) ($r$= 10.5 arcsec, PA=53\degr~, 
$\epsilon$=0.34).

$\epsilon$ and PA show strong variations at 18-20
arcsec from the nucleus, where the spiral arms begin (Fig. 10e) and
produce a small bump over the bulge+disk fit (Fig.  10f). The bar is
also apparent as a small enhancement over this fit.  However, due to
the relative weakness of the bar and arms, the residuals remain small
(Fig.  10g).
The J/K' image is quite smooth except in the very nucleus, which has
much redder colors (Fig.  10d) than the outer galaxy. 

\subsection{NGC~4785}

This galaxy has a bright thick bar (Fig. 11a) and a complex spiral
structure with several arms, including a thin arm to the south with
several blobs. Note that the major axis orientation varies with radius
in the central 14 arcsec (Fig. 11e).

The sharp-divided image reveals the presence of a ring and shows well
the spiral arms (Fig.  11b). These features are also clearly visible
in the difference image (Fig. 11c), where a small central bar seems to
be present. The existence of such a small bar is confirmed by the J/K
image (Fig. 11d) and by the variations of $\epsilon$ and PA with
radius (Fig. 11e) both in ground based and HST images.  The
sharp-dividing method applied to the NICMOS HST image shows an inner
elongation of $\approx$ 1.5 arcsec along PA=109\degr~ (Fig. 31).

The colour image is quite constant except at the location of the arms
and inner ring which appear redder in the J/K' color image (Figs. 11d
and 11h).

\subsection{NGC~5347}

The K' image of this galaxy only shows a large bar, the rest of the
structure being very smooth with hardly any hint for the presence of
spiral arms (Fig. 12a), in agreement with Mulchaey et al. (1997). Even
the sharp-divided image shows no structure (Fig. 12b) except for the
bright nucleus.

However, the extremities of the bar appear well in the difference
image (Fig. 12c), and the existence of a bar is confirmed by the PA
and $\epsilon$ variations both from ground based and HST images
(Fig. 12e) (in agreement with Mulchaey et al.  1997) and by the excess
over the bulge+disk fit (Figs. 12f and 12g).

Surprisingly, the J/K' image reveals a double structure separated by
$\approx$ 3 arcsec, resembling a double nucleus (Fig. 12h). The close
inspection of the color optical to near-infrared image by Regan \&
Mulchaey (1999), shows that this may result from the dust lane
crossing the nucleus.

\subsection{NGC~5728}

This galaxy is larger than the size of our infrared images, so we will
only present here the properties of the inner regions.

The K' image clearly shows the presence of a small bar within the
large bar (Fig. 13a), both already reported by Wozniak et al. (1995)
in their BVRI images. The small bar appears even more strongly in the
sharp-divided image (Fig. 13b), where a small inner ring is also visible 
($r \approx$ 4 arcsec, corresponding to the ring reported
by Buta \& Crocker 1993). The small bar and ring as well as the
extremities of the large bar are also visible in the difference image
(Fig. 13c), while the nucleus and ring show weaker J relative to K'
than the rest of the galaxy (Fig. 13d).

Two maxima of $\epsilon$ coupled with a constant PA (Fig. 13e)
allow to determine both primary and secondary bar parameters 
(see also Shaw et al. 1993, Wozniak et al. 1995).

Since we obviously do not reach the galaxy disk, the bulge+ disk fit
cannot be fully correct (Figs. 13f and 13g).

\subsection{ESO~139-12}

This galaxy shows a very regular structure in the K' band (Fig. 14a),
with a luminous bulge and no bar. No structure appears either in the
sharp-divided or difference images (Figs. 14b and 14c), and the J/K'
image appears to be very smooth (Figs. 14d and h). A faint and quite
curved dust lane is visible in the published HST image (Malkan et al.
1998) within the central 2 arcsec only.

While $\epsilon$ does not vary much with radius, the PA does vary
strongly as the isophotes are seen to rotate (Fig. 14e), probably due
to the effect of the flocculent spiral structure.

The bulge+disk fit is acceptable, with residuals smaller than 20\%
throughout (Figs. 14f and 14g).
The J/K' image shows a red nuclear region, also visible in the color
gradient plot (Fig.  14h).

\subsection{NGC~6814}

This galaxy shows a beautiful spiral structure on the K' image, with
spiral arms emerging from a thick bar (Fig. 15a). Our image appears to
be quite similar to that of Mulchaey et al.  (1997).

The spiral structure is seen in the sharp-divided image, where the bar
is traced as a faint elongation (Fig.  15b). The spiral structure is clearly
observed in the difference image (Fig. 15c).

Except for the very nucleus, the J/K' image remains constant
throughout the galaxy, excepting the innermost 2 arcsec, which are
redder (Figs. 15d and 15h).

The bar parameters from (Fig. 15e) are in agreement with Mulchaey et
al. (1997). Note the strong and sudden change of PA at the radius
where the spiral arms begin.  However the bulge+disk fit is very good
(Figs. 15f and 15g).  The NICMOS HST image is saturated in the
nucleus, so it cannot be used to gather information on the presence of
inner structures. The bar is seen as a thick elongated 10 arcsec
structure along the NS direction in Fig. 32.

\subsection{NGC~6860}

This galaxy has an asymmetrical bar along PA$\sim 10\degr$, more
extended to the north than to the south (Fig. 16a). The ring is not
apparent in the K' image, where we barely detect the beginning of a
spiral arm north of the bar.

The bar in the sharp-divided image appears clearly bent (Fig. 16b),
possibly due to the dynamical effects of a small inner bar (Figs. 16b,
16c, 16d). The difference image (Fig. 16c) also evidences a faint
spiral arm detaching from the southern end of the outer bar to the
north east, and a brighter, tighter arm starting at the west end of
the inner bar to the South. 
The variations of $\epsilon$ and PA with radius
(Fig.  16e) hints on the presence of a secondary bar.

The J/K' image confirms the existence of a small redder inner bar
(Fig. 16d), 
that is also evidenced by the dust lane structure in the HST image by
Malkan et al. (1998). The main bar and the beginning of the spiral
arms appear as bumps in the bulge+disk fit (Figs. 16f and 16g).  The
color gradient is very steep in the central regions (Fig. 16h).

\subsection{NGC~6890}

Despite its classification as a non-barred galaxy, the K' image of NGC
6890 reveals a strong bar (Figs. 17a and e), together with the beginning of
two spiral arms at the extremities of the bar, the northern arm being
brighter than the southern one (Fig. 17b). These features
are also apparent in the K' image by Mulchaey et al. (1997), but these
authors give PA=179\degr~ corresponding to a radius r $\approx$ 14
arcsec, where the PA strongly decreases. However, their plots of the
PA and $\epsilon$ show that the bar actually extends up to about 6
arcsec with PA $\approx$ 15\degr, in agreement with our values. We
also note that, even considering the differences in the photometry
(see Table \ref{compara}) the isophotal levels differ by more than 6
magnitudes, those by Mulchaey et al. being extremely bright and most
probably incorrect.

The difference image reveals a beautiful spiral structure as well as a
possible inner ring (10-15 arcsec) (Fig. 17c). This inner ring seems
confirmed by the J/K' image (Fig. 17d), where J appears to be somewhat
larger around the nucleus than in the very central region.  Notice a
smooth decrease of J relative to K' with increasing radius from 2
arcsec outwards (Fig. 17h). The NICMOS HST image shows spiraling
structure reaching the central arsecond, with some hints of dust at
0.5 arcsec (Fig. 33).

The bulge+disk fit is quite good except in the arm regions (Figs. 17f
and 17g).

\subsection{NGC~6951}

This is the first galaxy of our sample for which all the necessary
data has been acquired (imaging and spectroscopy); it has been studied
in detail by P\'erez et al. (1999) and will not be presented here. We
just note that no secondary bar is detected even in NIR HST images.

\section{Description of individual non active galaxies.}

\label{non-active}

\subsection{NGC~151}

This galaxy is larger than the size of our infrared images, so we only
present here the
properties of the inner regions.

The K' image shows a strong and broad bar, a disky structure, a large
ring, and a nuclear region roughly perpendicular to the bar
(Fig. 18a). Besides confirming the presence of a weak ring, the
sharp-divided image reveals the existence of a small bar roughly
perpendicular to the large bar (Fig. 18b). These features appear very
clearly in the difference image (Fig.  18c) and are quantified by
looking at the variations of $\epsilon$ and PA with radius
(Fig. 18e). The Pa$\alpha$ image by B\"oker et al. (1999) shows
that there is star formation ocurring at the ends of the bar and along
the ring. The sharp-masking method applied to the HST F160W image
confirms the presence of the inner bar which can be traced closer to
the nucleus ($r\approx$ 2.5 arcsec, PA=68\degr, Fig. 34).

The J/K' image shows a roughly circular redder region of 7 arcsec radius
(Fig.  18d); the inner bar is seen as a slightly
redder elongation along PA$\approx$ 90\degr.

The bulge+disk fits are merely indicative since the image is too small
to reach the disk (Fig. 18f). They show the existence of the two
bars. 

\subsection{IC~454}

This galaxy has a particularly large bar, with the west arm more
visible than the east one (Fig. 19a). The bright feature at the north
west end of the bar is most probably a superimposed star. The central
region of the bar and the west arm are clearly seen in the
sharp-divided image (Fig.  19b), while the extremities of the bar
appear more strongly on the difference image (Fig.  19c). This
illustrates the fact that the sharp-dividing method is very well
suited to detect elongated features which do not always appear as
clearly in the difference images (in the cases where they can be
nicely fit by ellipses).  The PA and $\epsilon$ plots evidence the
presence of a secondary bar (see also Fig. 19c).

The J/K' image and (J-K') color gradient appear quite constant except
in the bulge region and central bar where (J-K') is redder (Figs. 19d
and 19h).
The bulge+disk fit is quite good except at the extremities of the bar
(Figs. 19f and 19g).

\subsection{NGC~2712}

This galaxy also has a very large bar, and well developed arms, the
eastern being brighter than the western one (Fig. 20a). The
sharp-divided image (Fig. 20b) reveals the presence of a small bar
within the large bar. This is confirmed by the difference image
(Fig. 20c), where the spiral arms are also apparent. The $\epsilon$
and PA variations allow to quantify the parameters for both bars 
(Figs. 20e and 20h).

The J/K' image shows that the small bar is bluer, at variance with
the other nuclear regions which all appear redder in (J-K')
(Fig. 20e), including the surrounding emission. The joint analysis
with the optical data will allow a full interpretation of this
result. This structure is also visible in the color gradient, which
remains constant further out (Fig. 20h).

The bulge+disk fit is obviously not good because of the strong bar and
spiral arms (Figs. 20f and 20g).

\subsection{NGC~2811}

The entire galaxy almost fits in the K' image, which has a quite
regular aspect, with no obvious bar or arms (Fig. 21a). This image is
very similar to that by Jungwiert et al. (1997) in the H band.  On the
other hand, the sharp-divided and difference images clearly reveal the
presence of a bar, a smaller bar and weak spiral arms (Figs. 21b and
21c). The existence of two bars is confirmed by the variations of 
$\epsilon$ and PA (Fig.  21e).

The J/K' image is smooth, with a reddish bar and a blue surrounding ring
(Fig. 21d).
The bulge+disk fits are satisfactory, with small residuals throughout
(Figs. 21f and 21g).

\subsection{NGC~3571}

This galaxy does not fit entirely in the K' frame. The K' image shows
no evidence for a bar (Fig. 22a), but the sharp-divided image does
show the existence of a small thick bar (Fig. 22b). Due to its small
size and relative roundness, the bar cannot be detected either in the
difference image (Fig. 22c) nor in the $\epsilon$ and PA plots
(Fig. 22e).  However, the ellipse fitting to the HST image evidences
this structure as a clear elongation of $\sim$2 arcsec along
PA=90\degr (Fig. 22e). This is confirmed by the HST sharp-divided
image (Fig. 35) derived from the B\"oker et al. (1999) data.

The J/K' image is quite smooth, with the central region somewhat
redder than the outer zones (Fig. 22d).

The bulge+disk fits are quite good (Figs. 22e and 22f) even if we do
not have an image of the entire galaxy (Fig. 23h).

\subsection{NGC~3835 (UGC~6703)}

A thick bar and two spiral arms can be seen in the K' image (Fig. 23a)
and even better in the sharp-divided image (Fig. 23b), while the
difference image only shows the spiral arms clearly, but not the bar
(Fig. 23c).

The characteristics of the bar can be seen in the variations of 
$\epsilon$ and PA with radius (Fig. 23e).

The J/K' image shows a small extension (4 arcsec in radius) along the
east-west direction where the (J-K') colour is bluer than for
the rest of the galaxy (Fig. 23d). Better resolution images should be
obtained in order to ascertain what causes the misalignment between
this structure in the color image and the inner bar.

Because of the presence of the spiral arms, the bulge+disk fits are
not very good (Figs. 23f and 23g).

\subsection{NGC~4162 (UGC~7193)}

The K' image of this galaxy is quite irregular (Fig. 24a); two arms
with a flocculent appearance can be seen, but no bar. The
sharp-divided image does not show any strong feature (Fig. 24b). On
the other hand, the difference image reveals a beautiful spiral
structure (Fig. 24c), with a circumnuclear ring, three rather closed
spiral arms starting from the ring towards the north and three other
more open spiral arms also starting from the ring but towards the
south (Fig. 24c). 

The $\epsilon$ and PA variations with radius may reveal the presence
of a bar (Fig. 24e);
however, this feature does not appear very clearly in these plots and
we only have some hints of a small thick bar in Fig. 24b, so we cannot
be completely certain of its detection.

The bar and spiral arms appear in the bulge+disk fits in J and K'
as bumps at about 12, 20 and 38 arcsec (Figs. 24f and 24g).
The J/K' image is quite smooth, with (J-K') somewhat redder at the
center (Fig. 24d). 

\subsection{NGC~ 4290}

The K' image shows the bar, spiral arms and a large weak external ring
(Fig. 25a); the central isophotes are twisted relative to the bar. The
sharp-divided image seems to show a small structure in the center at
PA=90\degr~ (Fig. 25b); the bar and spiral arms are clearly
seen. These structures appear even more strongly in the difference
image (Fig. 25c). The  $\epsilon$ and PA variations are also
consistent with the existence of two bars 
(Fig.  25e); however, it is difficult to say if the small structure in
the center is really a small bar within the bar because it is thick
and faint.

The bulge+disk fits are good except in the bar and spiral arm regions
(Figs. 25f and 25g).  The J/K' image and the color gradient are very
smooth, with a slightly redder nucleus (Figs.  25d and 25h).

\subsection{NGC~4779}

A mosaic of images was obtained for this galaxy, so our data really
encompasses the entire object.

The K' image shows a strong bar, and weak flocculent and asymmetric
spiral structure (Fig. 26a), as confirmed by the sharp-divided image
(Fig. 26b). The bar and beginning of the spiral arms appear more
clearly on the difference image (Fig. 26c). Chapelon et al. (1999)
find a bar with PA=5\degr~ and $r\approx$ 16
arcsec. The bar that we measure is somewhat longer (Fig. 26e), 
in agreement with
Friedli et al.'s (1996) result that bars are generally longer in K
than in R (see also NGC 3660).

The J/K' colour image is fairly smooth (Fig. 26d). The bulge+disk fits
clearly show the bar and spiral arm regions (Figs. 26f and 26g). 

\subsection{NGC~6012 (UGC~10083)}

The K' image shows a bar with somewhat peculiar sharp edges; no
evidence either for spiral arms or for a ring is seen, in spite of the
classification of this galaxy as RSBR2* (Fig. 27a). This image is
consistent with the smaller one obtained by de Jong \& van der Kruit
(1994).  The peculiar aspect of the bar is confirmed by the
sharp-divided and difference images (Figs. 27b and 27c), where the bar
seems to be crossed by a dust lane. In this case, the bar would be
young and star formation should be observable along it; we will
therefore look for evidence for star formation in our spectra.

The difference image also shows evidence for bright spots towards the
edges of the bar, specially towards the north (Fig. 27c); although
this could be a star superimposed on the galaxy, the fact that there
is a faint southern counterpart tends to indicate that these regions
may both be in the galaxy.

The variations of $\epsilon$ and PA with radius (Fig. 27e) 
give bar parameters in agreement
with de Jong's (1996) results.

The bulge+disk model fits the profiles very nicely except in the
region of the bar and bright spots (Figs. 27f and 27g).
The J/K' image and color gradient are quite smooth throughout the galaxy
(Figs. 27d and 27h).

\subsection{NGC~6155}

A small bar is visible on the K' image, together with a spiral arm
starting northward and another one wrapped towards the south east
(Fig. 28a). Note that the center is displaced relatively to the
centroid of the outer isophotes.

The bar and north spiral arm appear faintly on the sharp-divided image
(Fig. 28b) and much more clearly on the difference image
(Fig. 28c). The variations of $\epsilon$ and PA with radius
(Fig. 28e) show that the bar reaches a radius of 6 arcsec 
(PA=120\degr). A larger bar may be present up to 15 arcsec 
(PA=160\degr); however, we cannot be certain that it is a bar since the
spiral arms seem to start at a smaller radial distance than the edges
of this structure.

The bulge+disk model shows the strong contribution of the bar and
spiral arms (Figs. 28f and 28g).

The J/K' image is fairly smooth and becomes redder at the 
very center (Figs.  28d and 28h). 

\section{Summary}\label{resume}

We have observed a sample of 29 isolated spiral galaxies: 18 host an
AGN (Seyfert 1 or Seyfert 2) and 11 are non-active galaxies. We
present here the infrared data in the J and K' bands, as well as the
image analysis. This includes sharp-divided images obtained by
dividing the observed images by their filtered counterparts,
difference images, obtained by fitting ellipses to the isophotes and
subtracting such models to the observed images, and colour J/K'
images. A bulge+disk model was fit to the image profiles, and the
corresponding fit parameters are given.

Four (one) out of five (one) of the optically classified non-barred
active (control) galaxies result to harbour a bar. Three of them had
already been described as barred galaxies, as derived either from NIR
(UGC 1395, which also has a secondary bar, and NGC 6890) or optical I
(NGC 3281) analyses. The other two (NGC 2639 and NGC 4162) are
classified as barred for the first time. For 15 (9 active, 6 control)
out of 24 (14 active, 10 control) of the optically classified barred
galaxies (SB or SX) we find that a secondary bar (or a disk, a lense
or an elongated ring; see table \ref{bars} for those cases for which the 
reported central elongation remains uncertain) is present.

A discussion on the physical properties of these galaxies, together
with the comparison of the properties of active versus non-active
galaxies derived from these data will be presented in a forthcoming
paper (M\'arquez et al., in preparation).

\begin{acknowledgements}

I.~M\'arquez acknowledges financial support from the Spanish Ministerio
de Educaci\'on y
Ciencia (EX94-8826734). This work is financed by DGICyT grants PB93-0139
and PR95-329. We
acknowledge financial support from INSU-CNRS for several observing
trips. Financial support
to develop the present investigation has been obtained by the
French-Spanish HF1996-0104 and
HF1998-0052 and the Chilean-Spanish bilateral agreement CSIC-CONICYT 99CL0018.

We thank Ron Probst, who made available the SQIID package for the
reduction of infrared
images within IRAF available to us. This research has made use of the
NASA/IPAC extragalactic
database (NED), which is operated by the Jet Propulsion Laboratory under
contract with the
National Aeronautics and Space Administration.

\end{acknowledgements}

\begin{table*}
\caption[ ]{Data for the sample galaxies in the RC3 catalogue. }
\begin{tabular} {lrrrlllrrrr}
\\
\hline
\\
Name      & $\alpha$(2000.0)& $\delta$(2000.0) &  PGC &   B  &  D$_{25}$ & R$_{25}$ & PA &  cz  &  t & Seyfert\\
  (1)  & (2)& (3) & (4)  & (5)  &  (6)   & (7)  & (8)  & (9)&(10) & (11)\\
\\
\hline
Active galaxies\\
\hline
\\
UGC  1395 & 01 55 22.2 & +06 36 41 &   7164 & 14.18 & 1.10 &  0.10 &   --- & 5184 & .SAT3..&1\\
IC    184 & 01 59 51.2 & $-$06 50 28 &   7554 & 14.66 & 1.02 &  0.31 &     7 & 5400 & .SBR1*.&2\\
IC   1816 & 02 31 51.3 & $-$36 40 17 &   9634 & 13.83 & 1.16 &  0.07 &   --- & 5086 & .SBR2P?&1\\
UGC  3223 & 04 59 09.4 & +04 58 31 &  16482 & 13.83 & 1.15 &  0.25 &    80 & 4723 & .SB.1..&1\\
NGC  2639 & 08 43 38.0 & +50 12 24 &  24506 & 12.59 & 1.26 &  0.22 &   140 & 3198 & RSAR1*&1\\
IC   2510 & 09 47 43.2 & $-$32 50 15 &  28147 & 14    & 1.10 &  0.27 &   148 & 2700 & .SBT2*.&2\\
NGC  3281 & 10 31 52.2 & $-$34 51 27 &  31090 & 12.60 & 1.52 &  0.30 &   140 & 3439 & .SAS2P*&2\\
NGC  3660 & 11 23 32.3 & $-$08 39 29 &  34980 & 15    & 1.43 &  0.09 &   115 & 3678 & .SBR4..&2\\
NGC  4253 & 12 18 26.4 & +29 48 48 &  39525 & 14.00 & 0.98 &  0.06 &   --- & 3819 & PSBS1*.&2\\
NGC  4507 & 12 35 37.1 & $-$39 54 31 &  41960 & 12.78 & 1.22 &  0.10 &   --- & 3499 & PSXT3..&2\\
NGC  4785 & 12 53 26.9 & $-$48 44 58 &  43791 & 13.21 & 1.29 &  0.31 &    81 & 3735 & PSBR3*.&2\\
NGC  5347 & 13 53 18.6 & +33 29 32 &  49342 & 13.16 & 1.23 &  0.10 &   130 & 2312 & PSBT2..&2\\
NGC  5728 & 14 42 24.1 & $-$17 15 12 &  52521 & 12.37 & 1.49 &  0.24 &     0 & 2885 & .SXR1*.&1\\
ESO139-12 & 17 37 39.5 & $-$59 56 29 &  60594 & 13.59 & 1.20 &  0.11 &    35 & 5200 & PSAT4P*&2\\
NGC  6814 & 19 42 40.7 & $-$10 19 25 &  63545 & 11.85 & 1.48 &  0.03 &   --- & 1509 & .SXT4..&1\\
NGC  6860 & 20 08 46.2 & $-$61 05 56 &  64166 & 13.68 & 1.13 &  0.25 &    34 & 4462 & PSBR3..&1\\
NGC  6890 & 20 18 17.9 & $-$44 48 25 &  64446 & 13.05 & 1.19 &  0.10 &   152 & 2471 & .SAT3..&2\\
NGC  6951 & 20 37 15.2 & +66 06 22 &  65086 & 11.91 & 1.59 &  0.08 &   170 & 1331 & .SXT4..&2\\
\\
\hline
Non-active galaxies\\
\hline
\\
NGC   151 & 00 34 02.8 & $-$09 42 20 &   2035 & 12.23 & 1.57 &  0.34  &   75 & 3654 & .SBR4.. &\\
IC    454 & 06 51 06.6 & +12 55 19 &  19725 & $-$   & 1.24 &  0.28  &  140 & 3945 & .SB.2.. &\\
NGC  2712 & 08 59 31.2 & +44 54 56 &  25248 & 12.38 & 1.46 &  0.26  &  178 & 1833 & .SBR3*. &\\
NGC  2811 & 09 16 11.3 & $-$16 18 47 &  26151 & 12.66 & 1.40 &  0.46  &   20 & 2514 & .SBT1.. &\\
NGC  3571 & 11 11 30.1 & $-$18 17 19 &  34028 & 12.99 & 1.48 &  0.46  &   94 & 3571 & PSXT1*. &\\
NGC  3835 & 11 44 05.6 & +60 07 13 &  36493 & 13.20 & 1.29 &  0.39  &   60 & 2452 & .SX.2*$^a$/ &\\
NGC  4162 & 12 11 51.8 & +24 07 24 &  38851 & 12.55 & 1.37 &  0.22  &  174 & 2542 & RSAT4.. &\\
NGC  4290 & 12 20 48.4 & +58 05 32 &  39859 & 12.66 & 1.37 &  0.16  &   90 & 3035 & .SBT2*. &\\
NGC  4779 & 12 53 50.8 & +09 42 33 &  43837 & 12.91 & 1.33 &  0.07  &   70 & 2793 & .SBT4.. &\\
NGC  6012 & 15 54 13.5 & +14 36 08 &  56334 & 12.96 & 1.32 &  0.14  &  168 & 1988 & RSBR2*. &\\
NGC  6155 & 16 26 08.4 & +48 21 58 &  58115 & 13.20 & 1.13 &  0.16  &  145 & 2429 & .SXT..$^b$ &\\
\hline
\\
\end{tabular}
\begin{footnotesize}

(1) Most common galaxy name; 

(2) and (3) $\alpha$ and $\delta$ coordinates (equinox 2000.0); 

(4) PGC number; 

(5) B magnitude; 

(6) isophotal diameter (at B=25 mag arcsec$^{-2})$, D$_{25}$, in units of
log(0.1$\times a$ (arcmin)); 

(7) axis ratio, R$_{25}$ in units of log$(a/b)$; 

(8) major axis position angle in degrees from N to E (when available); 

(9) velocity in km/s; 

(10) galaxy morphological type as given by the RC3,
except for $^a$ (.S..2*\ in the RC3) and $^b$ 
(.S?... in the RC3), for which the morphological type is from 
M\'arquez \& Moles (1996); 

(11) Seyfert activity class for active galaxies.

The data in columns 5-9 were taken from the RC3.\\
\end{footnotesize}
\protect\label{sample}
\end{table*}

\begin{table*}
\caption[ ]{Journal of Observations. }
\begin{scriptsize}
\begin{tabular} {llrcrll}
\\
\hline
\\
Name & Teles. & Obs. & Filter & Exp.  & Seeing & $\mu_{2\sigma}$ \\
     &        &date~~&        & time~ & FWHM   & (mag/ \\
     &        &      &        & (sec.)& (arcsec) & arcsec$^2$) \\
\\
\hline
\\
UGC  1395 & CAHA & 09/96 & K' & 5400 & 0.95& 20.7\\
IC    184 & CAHA & 09/96 & K' & 1800 & 0.95& 19.8\\
IC   1816 & LC   & 09/96 & K' & 3000 & -   & 20.1\\
          &      &       & J  &  800 & -   & 21.0\\
UGC  3223 & LC   &       & K' & 2700 & 1.2 & 19.9\\
NGC  2639$^{1}$ & TBL  & 03/97 & J  & 1000 & -   & 20.5\\
          &      &       & K' & 3000 & -   & 19.7\\
IC   2510$^*$ & LC   & 05/96 & K' & 3600 & 0.87& 19.6\\
          &      & 04/97 & J  & 1000 & 0.71& 20.8\\
NGC  3281 & LC   & 05/96 & K' & 3500 & 0.84& 19.8\\
          &      & 04/97 & J  & 1000 & 0.89& 21.1\\
NGC  3660 & LC   & 04/97 & J  & 1200 & -   & 21.3\\
          &      & 04/97 & K' & 2000 & -   & 20.2\\  
NGC  4253$^{3}$& TBL  & 03/97 & J  & 1000 & -   & 21.3\\
          &      &       & K' & 2800 & -   & 19.3\\
NGC  4507$^*$ & LC   & 05/96 & J  & 1400 & 0.80& 21.0\\ 
          &      &       & K' & 3600 & 0.97& 20.2\\
NGC  4785$^{3}$& LC   & 04/97 & J  & 1000 & 0.71& 20.8\\ 
          &      &       & K' & 3000 & 0.71& 20.2\\ 
NGC  5347$^*$ $^{2}$& TBL  & 03/97 & J  & 1000 & 1.4 & 21.3\\
          &      &       & K' & 3060 & 1.9 & 19.4\\
NGC  5728 & LC   & 04/97 & J  & 1000 & 0.88& 20.9\\
	  &      &       & K' & 2800 & 0.85& 20.2\\
ESO139-12 & LC   & 05/96 & J  & 1000 & 1.0 & 21.2\\
          &      &       & K' & 3400 & 0.97& 19.5\\
NGC  6814$^{3}$ & LC   & 05/96 & J  & 1000 & 1.2 & 20.9\\
          &      &       & K' & 3200 & 0.87& 19.3\\
NGC  6860$^{3}$ & LC   & 04/97 & J  & 1000 & 0.87& 20.9\\
          &      &       & K' & 2600 & 0.89& 20.3\\
          & ESO   & 05/96 & K' &  720 & 0.89& 17.5\\
NGC  6890 & LC   & 04/97 & J  & 1000 & 0.76& 20.7\\
          &      &       & K' & 2600 & 0.78& 20.1\\
NGC  6951  & NOT  & 09/96 & J  & 1760 & 1.3 & 18.8\\   
           &      &       & K' & 3520 & 1.3 & 19.5\\   
\hline
\\
NGC   151$^{1}$ & LC   & 09/96 & J  &  800 &  -   & 21.1\\
          &      &       & K' & 1800 &  -   & 19.9\\     
IC    454 & TBL  & 03/97 & J  & 1000 & 1.2 & 20.3\\
          &      &       & K' & 2800 & 1.25& 19.2\\ 
NGC  2712 & TBL  & 03/97 & J  & 1000 & 1.1 & 20.7\\
          &      &       & K' & 3200 & 1.2 & 20.1\\
NGC  2811$^*$ & LC   & 05/96 & K' & 3400 & 1.0 & 19.8\\
          &      & 04/97 & J  & 1000 & 1.19& 20.7\\
NGC  3571$^{1}$ & LC   &       & J  & 1000 & 0.84& 21.1\\
          &      &       & K' & 3000 & 0.67& 20.1\\
NGC  3835$^*$ & TBL  & 03/97 & J  & 1000 &  -  & 21.3\\
          &      &       & K' & 3000 &  -  & 19.7\\        
NGC  4162 & TBL  & 03/97 & J  &  800 & 1.6 & 21.3\\
          &      &       & K' & 3000 & 1.1 & 20.0\\   
NGC  4290 & TBL  & 03/97 & J  & 1200 & 1.1 & 21.0\\
          &      &       & K' & 3000 & 1.1 & 19.5\\
NGC  4779 & TBL  & 03/97 & J  & 4$\times$1000&1.1&20.9\\ 
          &      &       & K' & 4$\times$1000&1.1&19.3\\
NGC  6012 & TBL  & 03/97 & J  & 1000 & 1.2 & 21.0\\
          &      &       & K' & 3600 & 1.35& 19.7\\
NGC  6155 & TBL  & 03/97 & J  &  600 & 1.15& 21.0\\
          &      &       & K' & 3200 & 1.65& 19.5\\
\hline
\end{tabular}

CAHA: 3.5m telescope at Calar Alto (0.32''/pixel); 
LC: 2.5m telescope at Las Campanas (0.348''/pixel); 
TBL: 2m Bernard Lyot telescope at Pic du Midi (0.5''/pixel); 
ESO: 2.2m telescope at La Silla (0.278''/pixel); 
NOT: Nordic Optical Telescope at La Palma. (0.51''/pixel)\\

$^*$ Galaxies with non-flat remaining background.\\
$^{1,2,3}$ Galaxies with NICMOS HST archive images, 
(1) published by B\"oker et al. (1999); (2) published by Regan \& Mulchaey 
(1999); (3) unpublished.  

\end{scriptsize}
\protect\label{journal}
\end{table*}

\begin{table*}
\caption[]{Bulge+disk decomposition  parameters and magnitudes}
\label{decomposition}
\begin{scriptsize}
\begin{tabular} {lrrrrrrr|rrrrrrr|rrrr}
\\
\hline
\\
Galaxy   &$\mu_{d}$ &r$_{d}$  &m$_{d}$   &$\mu_{b}$ & r$_{b}$  &m$_{b}$  &m$_T$ &$\mu_{d}$ &r$_{d}$ &m$_{d}$  &$\mu_{b}$  &r$_{b}$  &m$_{b}$   &m$_T$  &m$_J^{ap}$ &r$_{J}^{ap}$ &m$_{K'}^{ap}$ &r$_{K'}^{ap}$\\
~              & J    &    &     &    &     &     &    &K' &    &     &      &     &      &       &     &   & &  \\
\\
(1)&(2)&(3)&(4)&(5)&(6)&(7)&(8)&(9)&(10)&(11)&(12)&(13)&(14)&(15)&(16)&(17)&(18)&(19)\\
\hline
\\
UGC 1395 & -     & -   & -    & -   & -    & -   & -  &19.8 &21  &10.5 &16.3  &1.17 &12.6 &10.4& -   &-   &10.82& 35\\
IC   184 & -     & -   & -    & -   & -    & -   & -  &18.2 &14  &9.9  &17.7  &4.9  &10.9 &9.5 & -   & -   &9.98& 35\\
IC  1816 &  18.8 &11   & 11.0 &11.2 & 0.2  &10.9 &10.2&18.1 &13  &9.8  &12.6  &0.54 &10.56&9.38&10.65& 45&9.52&45\\
UGC 3223 & -     & -   & -    & -   & -    & -   & -  &18.1 &20  &8.9  &15.9  &2.1  &10.9 &8.7 & -   & -   & 9.28&45\\
NGC 2639 &  17.6 &17   &  8.7 &18.2 & 9.6  & 9.9 & 8.4&16.7 &18  &7.8  &17.5  &14   &8.3  &7.3 &9.08 &50& 7.83&50\\
IC  2510 &  19 &19.2 & 10.1 &18.0 & 2.9  &12.3 &10. &17.9 &16  &9.1  &10.6  &0.3  &10.2 &8.8 &10.82&42& 9.56&42\\
NGC 3281 &  19.6 &92 & 7.03 &18.8 &15    &9.48 &6.9 &19.1 &109 &6.2  &17.8  &17   &8.3  &6.1 &9.42 &35& 8.43&35\\
NGC 3660 &  21.2 &44 & 10.3 &21.4 &57     &9.3  &8.9 &19.1 &18  &10   &19.8  &19   &10   &9.4 &10.63& 35&9.89&20\\
NGC 4253 &  18.4 &10   & 10.7 &17.3 & 2.7  &11.7 &10.3&17.0 &10  &9.4  &6.4   &0.1  &7.3  &7.2 &10.63& 50&9.21&50\\
NGC 4507 &  20.1 &31   &  9.9 &17.9 & 7.5  &10.1 &9.3 &18.9 &32  &8.7  &16.0  &4.8  &9.2  &8.2 &9.97& 44& 8.75&44\\
NGC 4785 &  17.1 &22   &  7.7 &12.3 & 0.7  &9.7  &7.6 &17.5 &25  &7.8  &15.0  &2.1  &9.9  &7.6 &9.73 & 28&8.77&28\\
NGC 5347 &  19.8 &23   & 10.3 &19.1 & 7.9  &11.2 &9.9 &18.2 &19  &9.1  &15.5  &2.2  &10.4 &8.8 &10.30& 48&9.00&48\\
NGC 5728 &  20.2 &103  &  7.4 &17.0 & 6.1  &9.6  &7.3 &18.1 &22  &8.8  &15.3  &3.6  &9.1  &8.2 &9.58&38 & 8.68&38\\
ESO139   &  20.0 &25   & 10.3 &15.6 & 1.3  &11.6 &10. &19.1 &26  &9.3  &15.2  &1.7  &10.6 &9.0 &10.54&45& 9.69&45\\
NGC 6814 &  19.4 &62   &  7.7 &18.5 & 9.2  &10.3 &7.6 &18.0 &40  &7.3  &16.6  &5.5  &9.5  &7.1 &9.30&47 & 8.04&47\\
NGC 6860 &  18.2 &11   & 10.3 &13.2 & 0.4  &11.5 &10. &18.1 &15  &9.6  &16.6  &3.9  &10.3 &9.1 &10.56&38& 9.49&38\\
NGC 6890 &  18.4 &16   &  9.7 &17.8 & 3.5  &11.7 &9.6 &17.2 &15  &8.6  &15.5  &1.8  &10.8 &8.5 &10.35&29& 8.97&38\\
NGC6951 &  20.4 &69   &  8.5 &19.4 & 22   &9.3  &8.1 &18.6 &55  &7.2  &17.4  &16 & 8.0 &6.8 &8.6&112 & 7.20&112\\
\hline
NGC  151 &  23.8 &191  &  9.7 &19.6 & 25   &9.3  &8.7 &19.3 &11  &11.4 &19.2  &15   &10.0 &9.7 &9.18 &42& 8.68&45\\
IC   454 &  19.4 &53   &  8.1 &18.2 & 5.8  &11.0 &8.0 &18.1 &74  &6.1  &16.6  &5.5  &9.5  &6.0 &10.02&45& 8.77&45\\
NGC 2712 &  19.0 &24   &  9.5 &10.2 & 0.3  &9.4  &8.7 &18.3 &26  &8.5  &13.5  &1.5  &9.3  &8.1 &9.69&55 & 8.59&55\\
NGC 2811 &  19.3 &47   &  8.2 &17.4 & 9.8  &9.0  &7.8 &17.4 &26  &7.6  &14.6  &3.6  &8.4  &7.2 &9.07 &38& 7.92&38\\
NGC 3571 &  20.1 &68   &  8.3 &18.4 & 9.0  &10.3 &8.1 &19.3 &50  &8.1  &17.3  &7.9  &9.4  &7.8 &10.09&38& 9.13&38\\
NGC 3835 &  18.8 &25   &  9.1 &16.9 & 2.7  &11.4 &8.9 &17.8 &23  &8.3  &18.4  &12 &9.7  &8.0 &10.22&55& 8.86&55\\
NGC 4162 &  18.7 &17   &  9.8 &17.3 & 2.7  &11.8 &9.6 &17.5 &18  &8.6  &13.7  &0.9  &10.5 &8.4 &9.78&55 & 8.64&55\\
NGC 4290 &  18.6 &20   &  9.4 &13.7 & 0.7  &11.2 &9.2 &17.6 &20  &8.3  &12.6  &0.7  &9.9  &8.1 &9.74 &60& 8.46&60\\
NGC 4779 &  20.4 &37   &  9.8 &16.0 & 1.8  &11.4 &9.6 &19.4 &37  &8.9  &14.8  &1.7  &10.2 &8.6 &10.06&65& 8.83&65\\
NGC 6012 &  19.1 &30   &  9.1 &16.6 & 0.8  &13.8 &9.0 &18.0 &30  &7.9  &11.8  &0.2  &11.6 &7.8 &10.00&48& 8.90&48\\
NGC 6155 &  19.2 &19   & 10.2 &20.3 &11.8  &11.5 &9.9 &18.5 &18  &9.5  &20.7  &52   &9.2  &8.6 &10.29&55& 9.19&55\\
\\
\hline
\\
\end{tabular}

(1) Galaxy name\\
(2) Equivalent surface brightness for the disk component in J band
\\
(3) Equivalent radius for the disk component in J band\\
(4) Computed total magnitude for the disk in J band\\
(5) Equivalent surface brightness for the bulge component in J band\\
(6) Equivalent radius for the bulge component in J band\\
(7) Computed total magnitude for the bulge in J band\\
(8) Computed total bulge $+$ disk J magnitude\\
(9)-(15) The same as (2)-(8) for the K' band\\
(16) J magnitude measured from simulated aperture photometry\\
(17) Radius of the circular aperture used to measure the J magnitude\\
(18) K' magnitude measured from simulated aperture photometry\\
(19) Radius of the circular aperture used to measure the K' magnitude\\
Surface brightnesses are given in magnitudes arcsec$^{-2}$ and radii in 
arcsec.

\end{scriptsize}

\end{table*}

\begin{table*}
\caption[ ]{Total magnitudes for the 9 galaxies with also K' and/or J
magnitudes from the litterature}
\begin{tabular} {lrrrr}
\\
\hline
\\
Galaxy  &K'$_{us}$& K'$_{others} (ref)$ &J$_{us}$& J$_{others} (ref)$ 
\\
\hline
\\
UGC  1395 & 10.82 & 10.3(2) & &  \\
NGC  2639 &  7.83 &  8.44(5) 	     & 9.08 & 9.40(5) \\
NGC  3281 &  8.43 & 9.12(1)          & 9.42 & 10.28(1) \\
NGC  4253 &  9.21 & 9.9(2),9.56(5),9.65(6) & 10.63& 11.04(5),11.01(6) \\
NGC  4507 &  8.75 & 9.28(1),8.9(4)   & 9.97 & 10.27(1)\\ 
NGC  5347 &  9.00 & 9.7(2),9.8(4)  & &  \\
NGC  5728 &  8.68 & 9.21(1)          & 9.58  &10.25(1) \\
NGC  6814 &  8.04 & 8.65(1),8.0(4)   & 9.30 & 9.72(1) \\
NGC  6890 &  8.97 & 9.26(1),8.0(4)   & 10.35 & 10.31(1) \\
NGC  6012 &  9.02 & 9.19(3)   & &  \\
\\
\hline
\\
\end{tabular}

(1) K magnitudes from Glass \& Moorwood 1985; (2) McLeod \& Rieke
1995; (3) de Jong 1996; (4) Alonso-Herrero et al. 1998; (5) Hunt et
al.  1999. (6) K' magnitudes from Mulchaey et al. 1997.

\label{compara}

\end{table*}

\begin{table*}
\caption[ ]{Parameters of the bar(s).}
\label{bars}
\begin{tabular} {lrlrlrlll}
\\
\hline
\\
Galaxy &Primary&Primary     & Primary & Detection$^*$  &Secondary& Secondary  & Secondary& Detection$^*$  \\
       &bar PA &bar         & bar size&           &bar PA   & bar        & bar size &           \\
       &       &$\epsilon$  &(arcsec) &           &         & $\epsilon$ &(arcsec)  &           \\
\\
\hline
\\
UGC  1395 & 145 & 0.56 &16 & f,o      &139 & 0.32 & 7 &s,h\\
IC    184 & 170 & 0.50 & 9 & i,f,d    &30  & 0.40 & 4 &f,h\\
IC   1816 & 110 & 0.50 &10 & i,s,f,d  &0:  & 0.20:& 1:&c,h\\
UGC  3223 &  75 & 0.71 &21 & i,s,f,d  &75: & 0.45:& 5:&f,h\\
NGC  2639 & 137 & 0.40 & 8 & f,s      & 117& 0.15 & 2 &s,h\\
IC   2510 & 145 & 0.61 &15 & i,s,f    &143:& 0.52:& 7:&s\\
NGC  3281 & 133 & 0.40 & 5 & i,s,o    &    &      &   &\\
NGC  3660 & 110 & 0.70 &23 & i,s,f,d  &    &      &   &\\
NGC  4253 & 105 & 0.50 &11 & i,s,f,d  & 5  & 0.10 & 2 &s,d,h\\
NGC  4507 &  52 & 0.35 & 9 & s,d,f,o  &    &      &   &\\
NGC  4785 &  65 & 0.50 &10 & i,f      & 82 & 0.35 & 5 &f,d,c,h\\
NGC  5347 & 105 & 0.60 &34 & i,d,f    &    &      &   &\\
NGC  5728 &  30 & 0.60 &44 & i,f,o    & 90 & 0.45 & 4 &i,s,f,d\\
ESO139-12 &     &      &   &          &    &      &   &\\
NGC  6814 &  25 & 0.30 &12 & i,s,f,o  &    &      &   &\\
NGC  6860 &  10 & 0.55 &10 & i,s,f    & 90 & 0.10 & 4 &f,d,s,c,h\\
NGC  6890 &  15 & 0.30 & 6 & s,f,o    &    &      &   &\\
NGC  6951 &  84 & 0.52 &45 & i,s,f,d,o&   &      &   &\\
\\
\hline
\\
NGC   151 & 155 & 0.45 & 18 & i,s,f,d &80 & 0.30 & 5 &s,f,d,h,c\\
IC    454 & 128 & 0.76 & 18 & i,s,f,d &133& 0.40 & 3 &s,f,c\\
NGC  2712 &  25 & 0.60 & 20 & i,d,f   & 5 & 0.25 & 5 &s,f,d,c\\
NGC  2811 &  30 & 0.50 & 20 & s,f     & 20& 0.30 & 5 &s,f,d\\
NGC  3571 &  90:& 0.25:& 2: & s,c,h,f &   &      &   &\\
NGC  3835 &  55 & 0.40 &  5 & s,f,c   &   &      &   &\\
NGC  4162 & 145:& 0.30:& 12:& f,s     &   &      &   &\\
NGC  4290 &  35 & 0.65 & 26 & i,s,f,d &55:& 0.35:& 6:&s,f,d\\
NGC  4779 &   8 & 0.65 & 30 & f,d     &   &      &   &\\
NGC  6012 & 155 & 0.65 & 16 & i,s,f,d &   &      &   &\\
NGC  6155 & 160:& 0.35:& 15:& i,f     &120& 0.40 & 6 &s,f,d\\
\\
\hline
\\
\end{tabular}

$^*$ We refer
to the image in which the bar is detected as following: 
i: direct image; s: sharp-dividing; 
f: ellipse fitting parameters; d: difference image (original $-$ model); 
c: color image; h: HST image; o: others (see text).\\

Notes: \\
All the elongated central
structures we call {\it secondary bars} 
could be either bars or elongated rings or disks;
kinematical data will help to discriminate among them.\\
The quantities followed by a column are those for which we are
not certain that such an elongation is detected.\\ 

\end{table*}

\newpage

\noindent
{\bf Figure captions}

Fig.1. (a) (Top-left) Image of UGC 1395 in the K' band. Contour
levels are 20.6, 20.4, 18.9, 18.6, 18.4, 17.8, 17.6, 17.1, 16.3 and
15.4 mag/arcsec$^2$. The horizontal scale bar corresponds to 10
arcsec; (b)(2nd-left) Sharp-divided K' image; (c) (3rd-left) Ellipse
model subtracted K' image; (d) (Bottom-left) (J-K') color image (not
available for UGC 1395); (e) (Top-right) PA (open circles) and
$\epsilon$ (solid triangles) {\it versus} ellipse major axis (in
arcsec) in K'.  The corresponding curves for J and HST images, when
available, are plotted as short-dashed lines and thick full lines
respectively; (f) (2nd-right) Surface brightness profiles in K' (open
circles) and J (solid triangles), when available.  Bulge and disk
components together with the bulge+disk profile are given as solid
lines; (g) (3rd-right) Residuals from the bulge+disk decomposition, in
percentage = $(F_{obs}-F_{fit})/F_{obs}$; symbols as in (f); (h)
(Bottom-right) (J-K') color gradient measured in circular rings,
calculated from the circular aperture curves of growth (not available
for UGC 1395).  As in all following figures, north is to the top and
east to the left. X-axis in Figs. (e) to (h) refers to the semi-major
axis (radius) in arcseconds.

Fig. 2. (a) (Top-left) Image of IC 184 in the K' band. Contour levels
are 19.3, 18.1, 17.5, 17.1, 16.8, 16.5, 16.3, 15.8 and 15.0
mag/arcsec$^2$.  (b), (c), (d) (not available), (e), (f), (g) and (h)
(not available) as in Fig.~1.

Fig. 3. (a) (Top-left) Image of IC 1816 in the K' band. Contour levels
are 19.5, 19.0, 18.7, 18.3, 18.0, 17.7, 17.3, 17.0, 16.7, 16.5, 16.0,
15.6 and 14.5 mag/arcsec$^2$. (b), (c), (d), (e), (f), (g) and (h) as
in Fig.~1.

Fig. 4.  (a) (Top-left) Image of UGC 3223 in the K' band. Contour
levels are 19.6, 18.6, 18.2, 17.9, 17.4, 17.1, 16.6, 16.2 and 15.5
mag/arcsec$^2$.  (b), (c), (d) (not available), (e), (f), (g) and (h)
(not available) as in Fig. 1.

Fig. 5.  (a) (Top-left) Image of NGC 2639 in the K' band. Contour
levels are 18.9, 18.0, 17.3, 16.4, 16.0, 15.6, 15.2, 14.8 and 14.2
mag/arcsec$^2$.  (b), (c), (d), (e), (f), (g) and (h) as in Fig.~1.

Fig. 6.  (a) (Top-left) Image of IC 2510 in the K' band. Contour
levels are 18.7, 18.3, 17.6, 17.2, 16.7, 16.3, 15.5 and 14.5
mag/arcsec$^2$.  (b), (c), (d), (e), (f), (g) and (h) as in Fig. 1.

Fig. 7. (a) (Top-left) Image of NGC 3281 in the K' band. Contour
levels are 18.5, 18.0, 17.5, 17.0, 16.8, 16.6, 16.4, 15.9, 15.4, 14.7
and 14.0 mag/arcsec$^2$. (b), (c), (d), (e), (f), (g) and (h) as in
Fig. 1.

Fig. 8. (a) (Top-left) Image of NGC 3660 in the K' band. Contour
levels are 18.6, 18.2, 17.4, 16.8, 16.4 and 15.9 mag/arcsec$^2$.  (b),
(c), (d), (e), (f), (g) and (h) as in Fig. 1.

Fig. 9. (a) (Top-left) Image of NGC 4253 in the K' band. Contour
levels are 18.7, 17.7, 17.3, 16.7, 16.4, 16.0, 15.2, 14.2 and 12.3
mag/arcsec$^2$.  (b), (c), (d), (e), (f), (g) and (h) as in Fig.~1.

Fig. 10. (a) (Top-left) Image of NGC 4507 in the K' band. Contour
levels are 18.9, 18.2, 17.7, 17.5, 17.2, 16.8, 16.1, 15.6 and 14.5
mag/arcsec$^2$.  (b), (c), (d), (e), (f), (g) and (h) as in Fig.~1.

Fig. 11. (a) (Top-left) Image of NGC 4785 in the K' band. Contour
levels are 18.4, 17.9, 17.4, 16.8, 16.4, 16.1, 15.9, 15.4 and 14.4
mag/arcsec$^2$.  (b), (c), (d), (e), (f), (g) and (h) as in Fig.~1.

Fig. 12. (a) (Top-left) Image of NGC 5347 in the K' band. Contour
levels are 19.0, 18.4, 17.7, 17.3, 16.5, 16.0 and 14.8 mag/arcsec$^2$.
(b), (c), (d), (e), (f), (g) and (h) as in Fig.~1.

Fig. 13. (a) (Top-left) Image of NGC 5728 in the K' band. Contour
levels are 18.3, 17.7, 17.0, 16.5, 16.0, 15.3, 14.8, 14.5 and 14.0
mag/arcsec$^2$.  (b), (c), (d), (e), (f), (g) and (h) as in Fig.~1.

Fig. 14. (a) (Top-left) Image of ESO 139-12 in the K' band. Contour
levels are 18.9, 18.4, 18.0, 17.7, 17.3, 16.9, 16.5, 16.0 and 14.5
mag/arcsec$^2$.  (b), (c), (d), (e), (f), (g) and (h) as in Fig.~1.

Fig. 15. (a) (Top-left) Image of NGC 6814 in the K' band. Contour
levels are 18.1, 17.7, 17.4, 17.2, 17.0, 16.7, 16.5, 16.2, 16.0, 15.6,
15.2 and 14.5 mag/arcsec$^2$. (b), (c), (d), (e), (f), (g) and (h) as
in Fig.~1.

Fig. 16. (a) (Top-left) Image of NGC 6860 in the K' band. Contour levels
are 19.0, 18.3, 17.7, 17.0, 16.5, 16.0, 15.1, 14.2 mag/arcsec$^2$.
(b), (c), (d), (e), (f), (g) and (h) as in Fig.~1.

Fig. 17. (a) (Top-left) Image of NGC 6890 in the K' band. Contour
levels are 19.1, 18.1, 17.3, 16.7, 16.4, 16.1, 15.7, 15.4, 14.8 and
14.0 mag/arcsec$^2$. (b), (c), (d), (e), (f), (g) and (h) as in
Fig.~1.

Fig. 18. (a) (Top-left) Image of NGC 151 in the K' band. Contour
levels are 18.3, 17.7, 17.0, 16.5, 16.0, 15.6, 15.1 and 14.6
mag/arcsec$^2$.  (b), (c), (d), (e), (f), (g) and (h) as in Fig.~1.

Fig. 19. (a) (Top-left) Image of IC 454 in the K' band. Contour levels
are 18.3, 17.6, 17.2, 16.9, 16.5, 16.2, 16.0 and 14.8 mag/arcsec$^2$.
(b), (c), (d), (e), (f), (g) and (h) as in Fig.~1.

Fig. 20. (a) (Top-left) Image of NGC 2712 in the K' band. Contour
levels are 18.5, 18.1, 17.5, 17.1, 16.6, 16.0 and 14.8 mag/arcsec$^2$.
(b), (c), (d), (e), (f), (g) and (h) as in Fig.~1.

Fig. 21. (a) (Top-left) Image of NGC 2811 in the K' band. Contour
levels are 18.0, 17.3, 16.8, 16.4, 16.1, 15.9, 15.6, 15.3, 14.7 and
14.0 mag/arcsec$^2$. (b), (c), (d), (e), (f), (g) and (h) as in
Fig.~1.

Fig. 22. (a) (Top-left) Image of NGC 3571 in the K' band. Contour levels
are 18.6, 17.7, 17.2, 16.6, 16.3, 16.0, 15.7, 15.4, 14.8 and 14.0 
mag/arcsec$^2$. (b), (c), (d), (e), (f), (g) and (h) as in Fig.~1.

Fig. 23. (a) (Top-left) Image of NGC 3835 in the K' band. Contour levels
are 19.4, 18.6, 18.1, 17.8, 17.3, 17.0, 16.6, 16.2, 16.0, 15.2 and 18.8 
mag/arcsec$^2$. (b), (c), (d), (e), (f), (g) and (h) as in Fig.~1.

Fig. 24. (a) (Top-left) Image of NGC 4162 in the K' band. Contour
levels are 19.1, 18.4, 18.0, 17.7, 17.3, 17.0, 16.5, 16.2, 15.8, 15.2
and 14.8 mag/arcsec$^2$. (b), (c), (d), (e), (f), (g) and (h) as in
Fig.~1.

Fig. 25. (a) (Top-left) Image of NGC 4290 in the K' band. Contour
levels are 18.7, 18.1, 17.7, 17.3, 17.0, 16.6, 16.2, 15.8, 15.2 and
14.8 mag/arcsec$^2$. (b), (c), (d), (e), (f), (g) and (h) as in
Fig.~1.

Fig. 26. (a) (Top-left) Image of NGC 4779 in the K' band. Contour levels
are 18.9, 18.1, 17.7, 17.3, 17.0, 16.7, 16.2, 15.8 and 14.8 mag/arcsec$^2$.
(b), (c), (d), (e), (f), (g) and (h) as in Fig.~1.

Fig. 27. (a) (Top-left) Image of NGC 6012 in the K' band. Contour
levels are 18.9, 18.5, 18.0, 17.7, 17.3, 17.0, 16.6, 16.2 and 15.8
mag/arcsec$^2$.  (b), (c), (d), (e), (f), (g) and (h) as in Fig.~1.

Fig. 28. (a) (Top-left) Image of NGC 6155 in the K' band. Contour
levels are 19.4, 18.7, 18.0, 17.7, 17.3, 17.0, 16.6, 16.2 and 15.8
mag/arcsec$^2$.  (b), (c), (d), (e), (f), (g) and (h) as in Fig.~1.

Fig. 29. Sharp-divided HST NICMOS image of NGC 2639.  The orientation
is given by square angle where the North is marked. The length of both
sides is 2''.

Fig. 30. Sharp-divided HST NICMOS image of NGC 4253. 
Scale and orientation are traced as in Fig. 29.

Fig. 31. Sharp-divided HST NICMOS image of NGC 4785.
Scale and orientation are traced as in Fig. 29.

Fig. 32. Sharp-divided HST NICMOS image of NGC 6814.
Scale and orientation are traced as in Fig. 29.

Fig. 33. Sharp-divided HST NICMOS image of NGC 6890.
Scale and orientation are traced as in Fig. 29.

Fig. 34. Sharp-divided HST NICMOS image of NGC 151.
Scale and orientation are traced as in Fig. 29.

Fig. 35. Sharp-divided HST NICMOS image of NGC 3571.
Scale and orientation are traced as in Fig. 29.

\end{document}